\documentclass[12pt,english]{article}
\usepackage{amsmath}
\usepackage{amssymb}
\usepackage{mathrsfs}
\usepackage{graphicx}
\usepackage{fullpage}

\makeatletter

\catcode`@=12 \topmargin 0 in \evensidemargin 0.0in \oddsidemargin
0.0in \textheight 8.5in \textwidth 6.5in
\usepackage{babel}
\makeatother
\newcommand {\be}{\begin{equation}}
\newcommand {\ee}{\end{equation}}
\newcommand {\ba}{\begin{eqnarray}}
\newcommand {\ea}{\end{eqnarray}}

\begin{document}
\thispagestyle{empty}
\begin{flushright}
IPM/P-2009/055\\
\today
\end{flushright}

\mbox{} \vspace{0.75in}

\begin{center}\textbf{\large  On the Oscillation of Neutrinos Produced by the
Annihilation of Dark Matter inside the Sun}\\

 \vspace{0.5in} \textbf{$\textrm{Arman Esmaili}^{\dag\S}$\footnote{arman@mail.ipm.ir} and $\textrm{Yasaman Farzan}^{\S}$\footnote{yasaman@theory.ipm.ac.ir}}\\
 \vspace{0.2in} \textsl{${}^\dag$Department of Physics, Sharif University of Technology\\P.O.Box 11365-8639, Tehran, IRAN}\\
 \vspace{0.2in}\textsl{${}^\S$School of Physics, Institute for Research in Fundamental Sciences (IPM)\\P.O.Box 19395-5531, Tehran, IRAN}\\
 \vspace{.75in}\end{center}

\baselineskip 20pt
\begin{abstract}
The annihilation of dark matter particles captured by the Sun can
lead to a neutrino flux observable in neutrino detectors.
Considering the fact that these dark matter particles are
non-relativistic, if a pair of dark matter annihilates to a
neutrino pair, the spectrum of neutrinos will be monochromatic. We
show that in this case, even after averaging over production point
inside the Sun, the oscillatory terms of the oscillation
probability do not average to zero. This leads to interesting
observable features in the annual variation of the number of muon
track events. We show that smearing of the spectrum due to thermal
distribution of dark matter inside the Sun is too small to wash
out this variation. We point out the possibility of studying the
initial flavor composition of neutrinos produced by the
annihilation of dark matter particles via measuring the annual
variation of the number of $\mu$-track events in neutrino
telescopes.
\end{abstract}

\section{Introduction}
Growing evidence from a wide range of cosmological and
astrophysical observations shows that about 82~\% of the matter
content of the universe is composed of Dark Matter (DM) whose
exact identity is yet unknown ({\it i.e.,}
$\rho_{DM}/(\rho_{DM}+\rho_{baryon})=82~\%$ \cite{pdg}). In the
literature \cite{Hooper:2009zm}, various candidates for DM have
been suggested among which Weakly Interacting Massive Particles
(WIMPs) are arguably the most popular class of candidates
\cite{WIMP}. The WIMPs are expected to propagate in the space
between the stars and planets just like an asteroid which is
subject to gravitational force from the various astrophysical
bodies. In time a considerable number of WIMPs will interact with
the nuclei inside the Sun and lose energy. If the velocity drops
below the escape velocity, the particle will be captured by the
gravitational potential of the Sun \cite{suncapture}. The DM
particles will eventually thermalize inside the Sun. Having a
relatively large density inside the core, the DM particles will
annihilate with each other. Depending on the annihilation modes,
two classes of the WIMP models can be identified: (1) DM pair
directly annihilate to neutrino pairs (that is $\nu \bar{\nu}$ or
$\nu \nu+\bar{\nu}\bar{\nu}$). (2) DM pair particles annihilate to
various particles whose{ subsequent decay produce}
neutrinos alongside other particles. In both cases, the neutrinos
can be detected in the neutrino telescopes such as IceCube,
provided that the cross section of the WIMPs with nuclei is of the
order of $\sim 10^{-6}$~pb or larger. Indirect detection of DM
through registering the neutrinos has been thoroughly studied in
the literature \cite{sunneutrino}.

The oscillation length of the neutrinos due to 12-mixing can be
estimated as
\begin{equation} L_{osc}={4\pi E_\nu \over \Delta m_{12}^2}\sim
3 \times 10^{11}~{\rm cm}~\left({E_\nu\over 100~{\rm
GeV}}\right)\left(\frac{8\times 10^{-5}~{\rm eV}^2}{\Delta
m_{12}^2}\right),\end{equation} which is of order of few percent
of the distance between the Sun and the Earth. If the energy
resolution of the detector ($\delta E/E$) is worse than 1~\% and
the width of the spectrum is larger than $\delta E/E$, averaging
the oscillatory terms is justified. However, for a monochromatic
spectrum, dropping the oscillatory terms may lead to an error.
Investigating the observable effects of these terms and
information that they carry is the subject of the present letter.
We shall demonstrate this effect by diagrams based on explicit
calculation. We also demonstrate that integrating over production
point does not justify dropping the oscillatory terms.

The DM particles captured inside the Sun are non-relativistic and
have an average velocity of $(3 T_\odot/m_{DM})^{1/2}\simeq 60
~{\rm km/sec}$. As a result, in the case that the DM pairs
directly annihilate to neutrino pairs, the spectrum of neutrinos
will be monochromatic. Considering the fact that $L_{osc}$ is of
the order of the variation of Earth and Sun distance over a year,
we expect these oscillatory terms lead to a significant variation
of the number of events during a year. This expectation is similar
to prediction of seasonal variation of beryllium line in the
seminal paper by Gribov and Pontecorvo \cite{gribov}.

In the papers of reference \cite{weiler}, dropping the oscillatory
terms, the oscillation of the neutrinos on the way to the detectors
has been studied. In papers \cite{mono,mono1,Blennow:2007tw}
propagation of the monochromatic neutrinos has been numerically
studied within the 3-$\nu$ oscillation scenario; but,
no emphasis has been put on the potential effects of the
oscillatory terms. In \cite{Blennow:2007tw} the potential effects
of the oscillatory terms and the seasonal variations have been
pointed out but since the emphasis was on the flavor blind initial
composition, the effects had not been explored.

In this paper we focus on the effects of oscillatory terms on the
flux of monochromatic neutrinos from the Sun without assuming a
democratic initial flavor content. Monochromatic flavor dependent
neutrino flux can emerge within various models. A particular
scenario leading to monochromatic neutrino flux has been recently
worked out in \cite{Farzan:2009ji}. The scenario is based on an
effective coupling of form $g_\alpha N \phi \nu_\alpha$ where $N$
is a Majorana neutrino and $\phi$ is the scalar playing the role
of the dark matter. Within this scenario, the main annihilation
mode is ${\rm DM}+{\rm DM}\to
\nu_\alpha+\nu_\beta,\bar{\nu}_\alpha+\bar{\nu}_\beta$ and the
flavor composition can be determined by the flavor structure of
the coupling $g_\alpha$. The scenario can be embedded within
various models that respect the electroweak symmetry
\cite{models}. Another example of a model that can give rise to
monochromatic neutrino spectrum with non-democratic flavor ratio
can be found in \cite{Allahverdi}. We evaluate the magnitude of
seasonal variation and discuss the condition under which such an
amount of variation can be in practice established. We propose
using the seasonal variation measurement as a tool to probe the
physics of DM annihilation.

In section~\ref{osc}, taking into account the matter effects
inside the Sun, we derive the minimum value of the width of the
spectrum leading the oscillatory terms to average to zero. We also
show that uncertainty in the production point will not lead to
vanishing of the oscillatory terms. In section~\ref{widthSpec}, we
investigate various sources that lead to widening of the spectrum
of monochromatic neutrinos. We then show that this widening is too
small to make dropping of the oscillatory terms justifiable. In
section~\ref{seasonal}, we demonstrate the observable effects of
the oscillatory terms in seasonal variation of the flux. In
section~\ref{dis}, we anticipate  the conclusions that can be
reached based on different observational outcomes. The results are
summarized in section~\ref{Con}.

\section{Effects of the oscillating terms \label{osc}}

In case that the DM pair annihilates to a neutrino pair and the
physics governing dark matter annihilation is flavor blind,
neutrinos of all three flavors can be produced in equal numbers.
In this case, the neutrino oscillation cannot alter the flavor
composition: if $F^0_e=F^0_\mu=F^0_\tau$, $\sum_\alpha P_{\alpha
\beta} F^0_\alpha =F^0_\beta$. However, DM interaction can in
general be flavor sensitive so neutrinos of different flavors can
be produced with different amounts. For example, it has been
recently suggested that there might be a correlation between
flavor structure of DM couplings and the neutrino mass matrix
\cite{Farzan:2009ji}. In fact, it is possible that a coherent
mixture of different flavors are produced at the source which
means that the density matrix can be non-diagonal in the flavor
basis. The density matrix, being a Hermitian matrix, can always be
diagonalized. Let us denote the eigenstates of the density matrix
at the source with $|\nu_\alpha\rangle$ which are in general
linear compositions of $|\nu_e\rangle$, $|\nu_\mu\rangle$ and
$|\nu_\tau\rangle$.

\begin{figure}[t!]
  \begin{center}
 \centerline{\includegraphics[bb=250 40 590
 570,keepaspectratio=true,clip=true,angle=-90,scale=0.47]{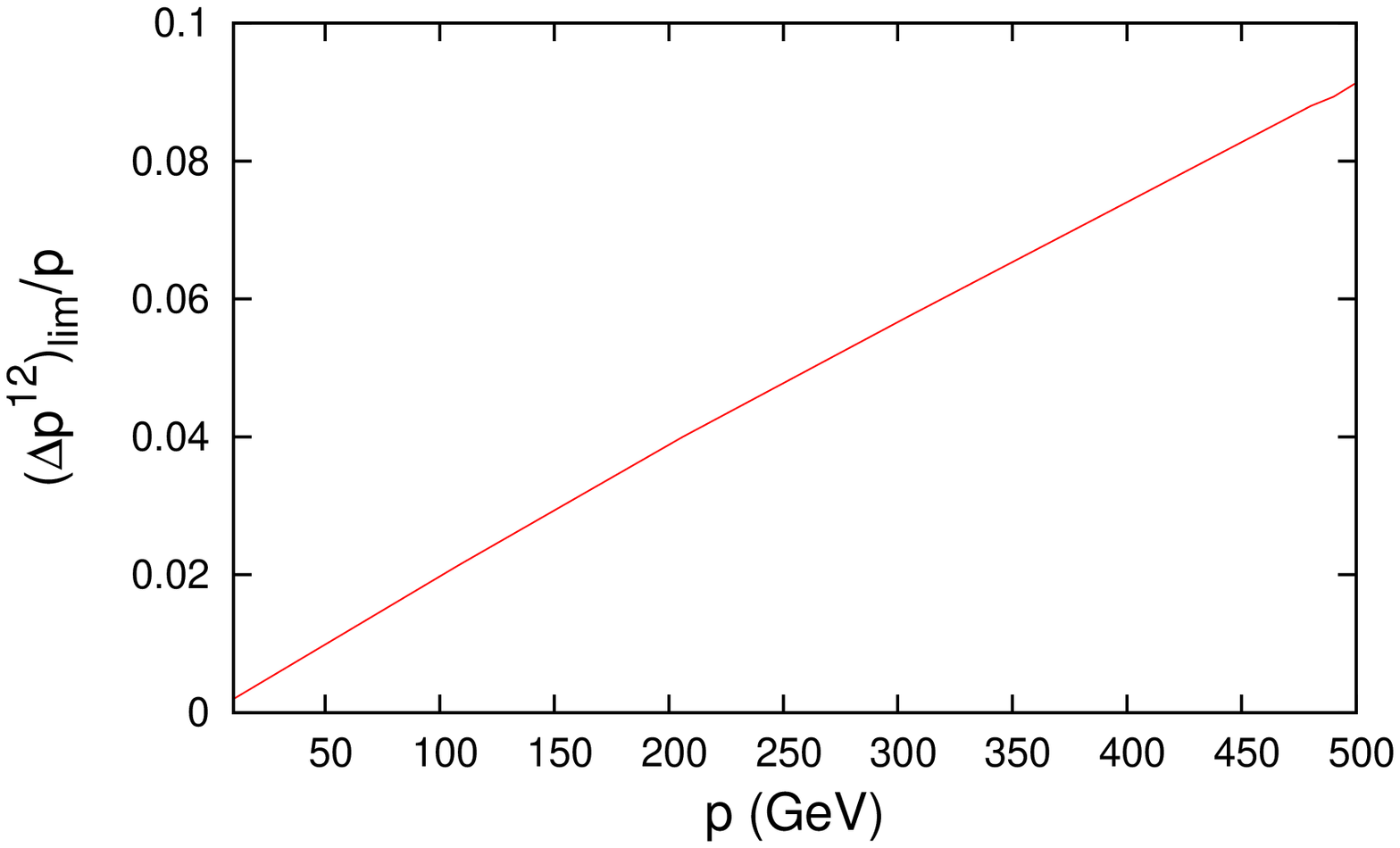}\includegraphics[bb=250 40 590
 570,keepaspectratio=true,clip=true,angle=-90,scale=0.47]{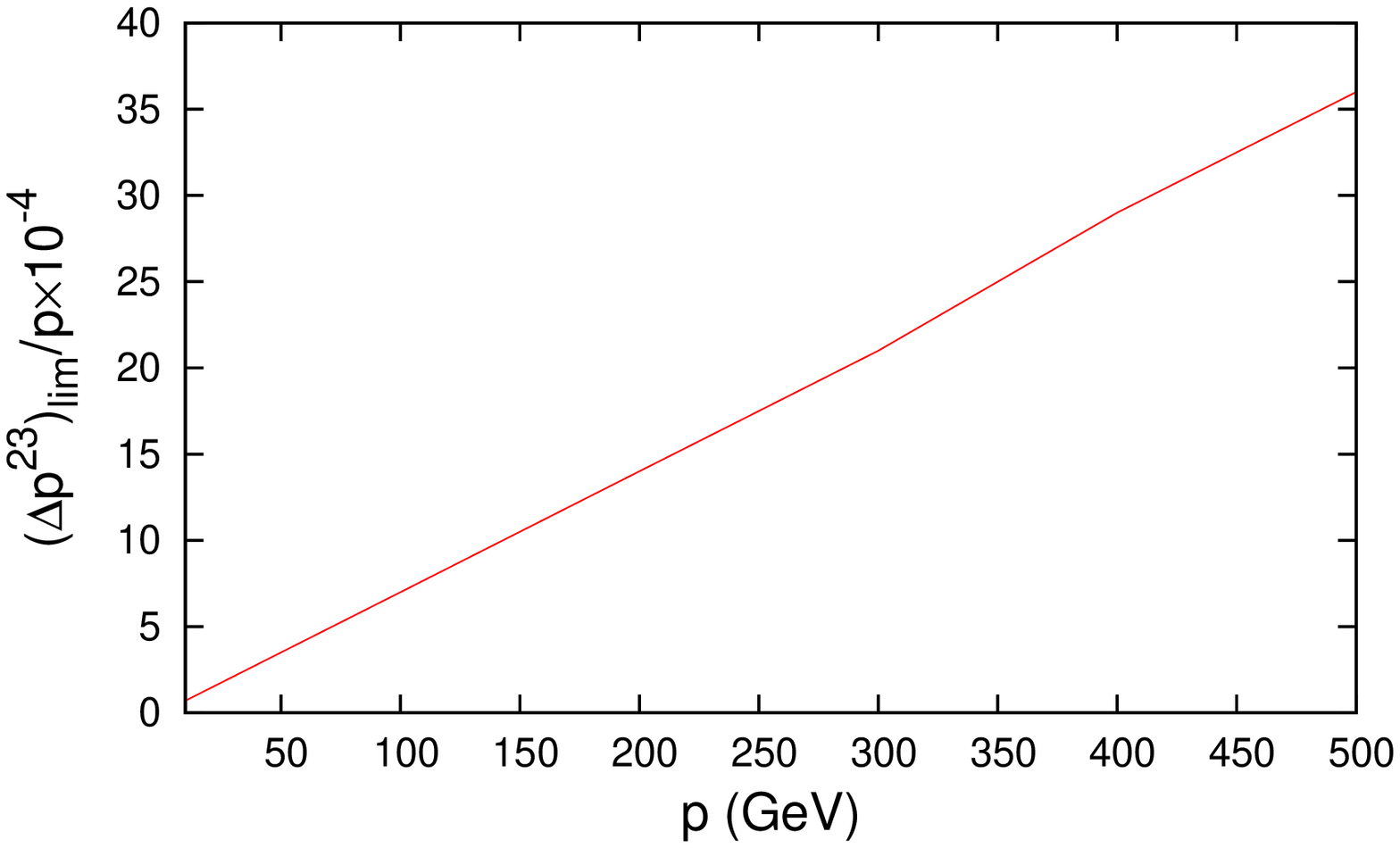}}
 \centerline{\hspace{0.45cm}(a)\hspace{8.45cm}(b)}
 \centerline{\vspace{-2.0cm}}
 \end{center}
 \caption{{\small The dependence of $(\Delta
p^{12})_{lim}/p$ and $(\Delta p^{13})_{lim}/p$ defined in
Eq.~(\ref{deltaplim}) on the momentum of neutrinos. 
The behavior shown in these graphs does not change by varying
$\theta_{13}$ in the allowed range $\theta_{13}<10\,^\circ$ {or
changing the flavor composition}. } }
  \label{fig1}
\end{figure}

Let us consider such a state with momentum $p$ produced inside the
center of the Sun. This state after propagating a distance $L$
evolves into \be \label{NuEvolve}|\nu_\alpha;p;L\rangle=a_{\alpha
1}(L)|1;p\rangle+a_{\alpha 2}(L)|2;p \rangle +a_{\alpha
3}(L)|3;p\rangle\ee where $|i;p\rangle$ is the neutrino mass
eigenstate (in the vacuum). Let us define $(\Delta p^{ij})_{lim}$
as the minimum value of the energy interval for which
\begin{equation}\label{deltaplim}\arg\left[{\left[a_{\beta
i}(0)(a_{\beta j}(0))^* (a_{\alpha i}(L/c))^* a_{\alpha
j}(L/c)\right]|_{p+(\Delta p^{ij})_{lim}} \over \left[a_{\beta
i}(0)(a_{\beta j}(0))^* (a_{\alpha i}(L/c))^* a_{\alpha
j}(L/c)\right]|_{p}}\right]=2\pi\ ,\end{equation} where $\alpha
\ne \beta$ and $L$ is the distance between the Sun and the Earth.
Suppose that the spectrum of the neutrino flux is almost
monochromatic with width $\Delta p \ll p$. If the width of the
spectrum is smaller than $(\Delta p^{ij})_{lim}$, the average of
the oscillatory terms will in general be nonzero. In this case, if
the energy resolution of detector, $\delta E$, is larger than
$\Delta p$, the corresponding oscillatory terms will lead to an
error in deriving the initial flux. Let us evaluate the numerical
value of $(\Delta p^{ij})_{lim}$. In vacuum, $(\Delta
p^{ij})_{lim}$ can be estimated as \be \frac{(\Delta
p^{ij})_{lim}}{p}=\frac{2\pi p}{L \Delta m_{ij}^2},\ee so taking
$L=1.5\times 10^{13}~{\rm cm}$ we find $(\Delta
p^{12})_{lim}/p=0.01~ (p/100~{\rm GeV})$ and $(\Delta
p^{13})_{lim}/p=0.0005~ (p/100~{\rm GeV})$. Notice that this
result is independent of  $\alpha$ and $\beta$. Of course, the
neutrinos produced in the center of the Sun have to pass through
the matter inside the Sun before reaching the detectors.
Figs.~(\ref{fig1}-a) and (\ref{fig1}-b) show $(\Delta
p^{12})_{lim}/p$ and $(\Delta p^{13})_{lim}/p$ versus the neutrino
momentum, taking into account the matter effects for various
values of $\theta_{13}$. To draw these figures, we have
numerically obtained the evolution of the neutrino states of
different flavors. Considering that the density of matter is
sizeable only inside the Sun with radius $R_\odot\sim 7 \times
10^{10}~{\rm cm} \sim L_{osc}$, we expect the numerical values of
$(\Delta p)_{lim}/p$ in the presence of matter to be almost
similar to that in vacuum. In fact, the results shown in
Fig.~\ref{fig1} fulfill this expectation: The order of magnitude
of $(\Delta p^{ij})_{lim}/p$ is the same as that in vacuum and
moreover independence of $(\Delta p^{ij})_{lim}/p$ from flavor
({\it i.e.,} $\alpha$ and $\beta$) is maintained.

Similarly to Eq.~(\ref{NuEvolve}), we can write \be
\label{BarNuEvolve}|\bar{\nu}_\alpha;p;L\rangle=\bar a_{\alpha
1}(L)|\bar 1;p\rangle+\bar a_{\alpha 2}(L)|\bar 2;p \rangle +\bar
a_{\alpha 3}(L)|\bar 3;p\rangle,\ee where $|\bar i;p\rangle$ is
the antineutrino mass eigenstate (in the vacuum). Due to matter
effects inside the Sun, even in the absence of the CP-violating
phase $\delta$, $a_{\alpha i}(L)\ne \bar{a}_{\alpha i}(L)$.
However since averaging of the oscillatory terms is mainly due to
the propagation of neutrinos and antineutrinos in the large empty
space between the Sun surface and the Earth, the value of $(\Delta
p^{ij})_{lim}/p$ for neutrinos and antineutrinos will be similar.
Direct numerical calculations confirm this claim.

Another effect that may lead to averaging of the oscillatory terms
is the difference in the production point; { \it i.e.,} the
difference in baseline. The dark matter particles are distributed
in a radius of \be \label{rDM}r_{DM}\approx\left({9 T\over  8\pi
G_N\rho m_{DM}}\right)^{1/2},\ee where $T$ and $\rho$ are
respectively the temperature and density inside the Sun center.
$G_N$ is the Newton gravitational constant. Inserting the
numerical values, we obtain $r_{DM}=2\times 10^8~{\rm cm}(100~{\rm
GeV}/m_{DM})^{1/2}=0.003R_\odot(100~{\rm GeV}/m_{DM})^{1/2}$.
Inside the volume around the Sun center with $r<r_{DM}$, the
density is high such that $V_e=\sqrt{2}G_F N_e\gg \Delta
m_{12}^2/p$, so $\nu_e$ and $ \bar{\nu}_e$ in practice correspond
to energy eigenstates. As a result, $\stackrel{(-)}{\nu_e}$ within
this volume does not go through oscillation. On the other hand,
the oscillation length corresponding to the 2-3 splitting, $4\pi
p/\Delta m_{atm}^2\sim 10^{10}~{\rm cm}$ is much larger than
$r_{DM}$ so averaging out the corresponding oscillatory terms is
not justified.
\begin{figure}[t!]
  \begin{center}
 \centerline{\includegraphics[bb=250 40 590
 570,keepaspectratio=true,clip=true,angle=-90,scale=0.48]{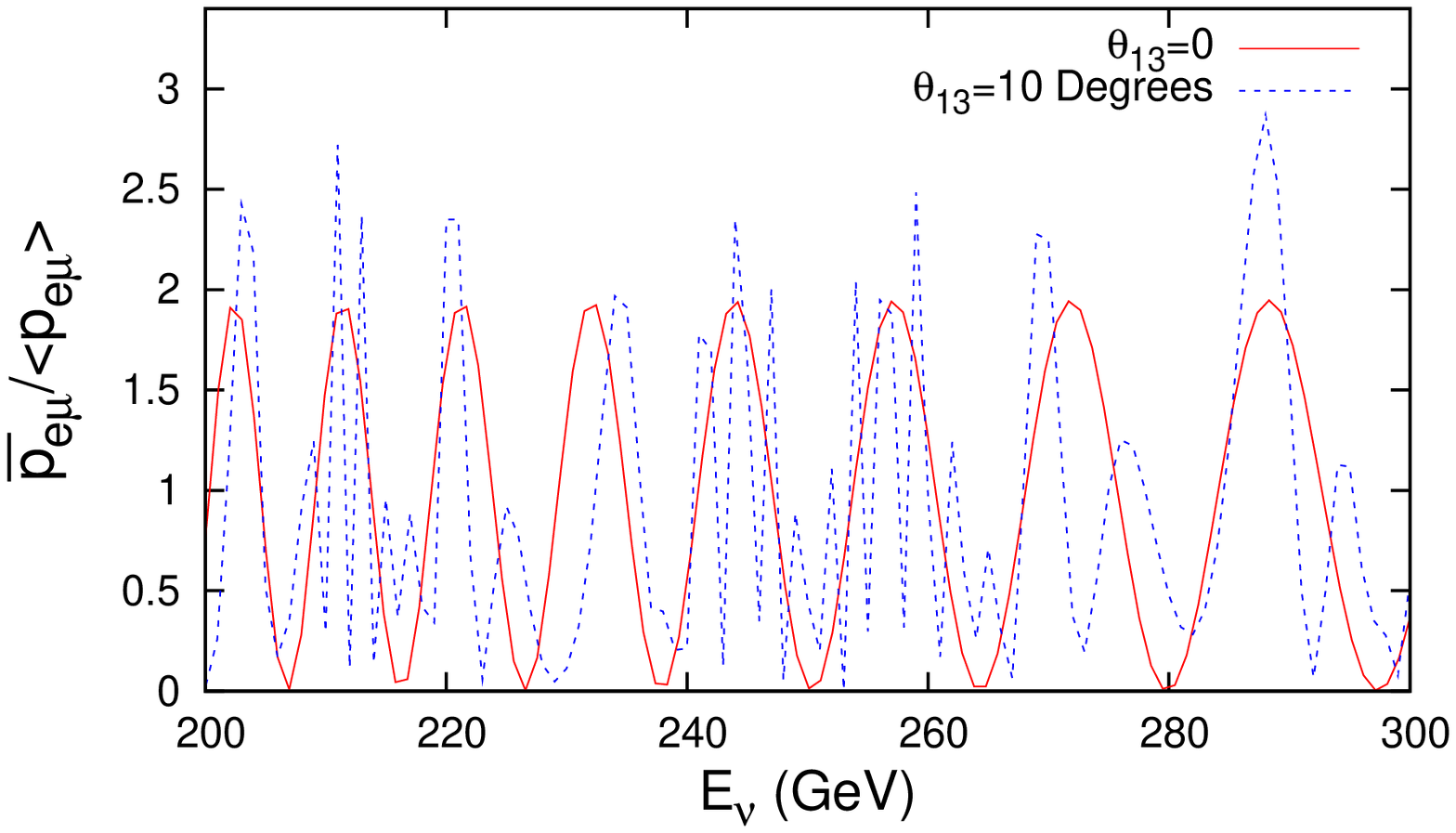}\includegraphics[bb=250 40 590
 570,keepaspectratio=true,clip=true,angle=-90,scale=0.48]{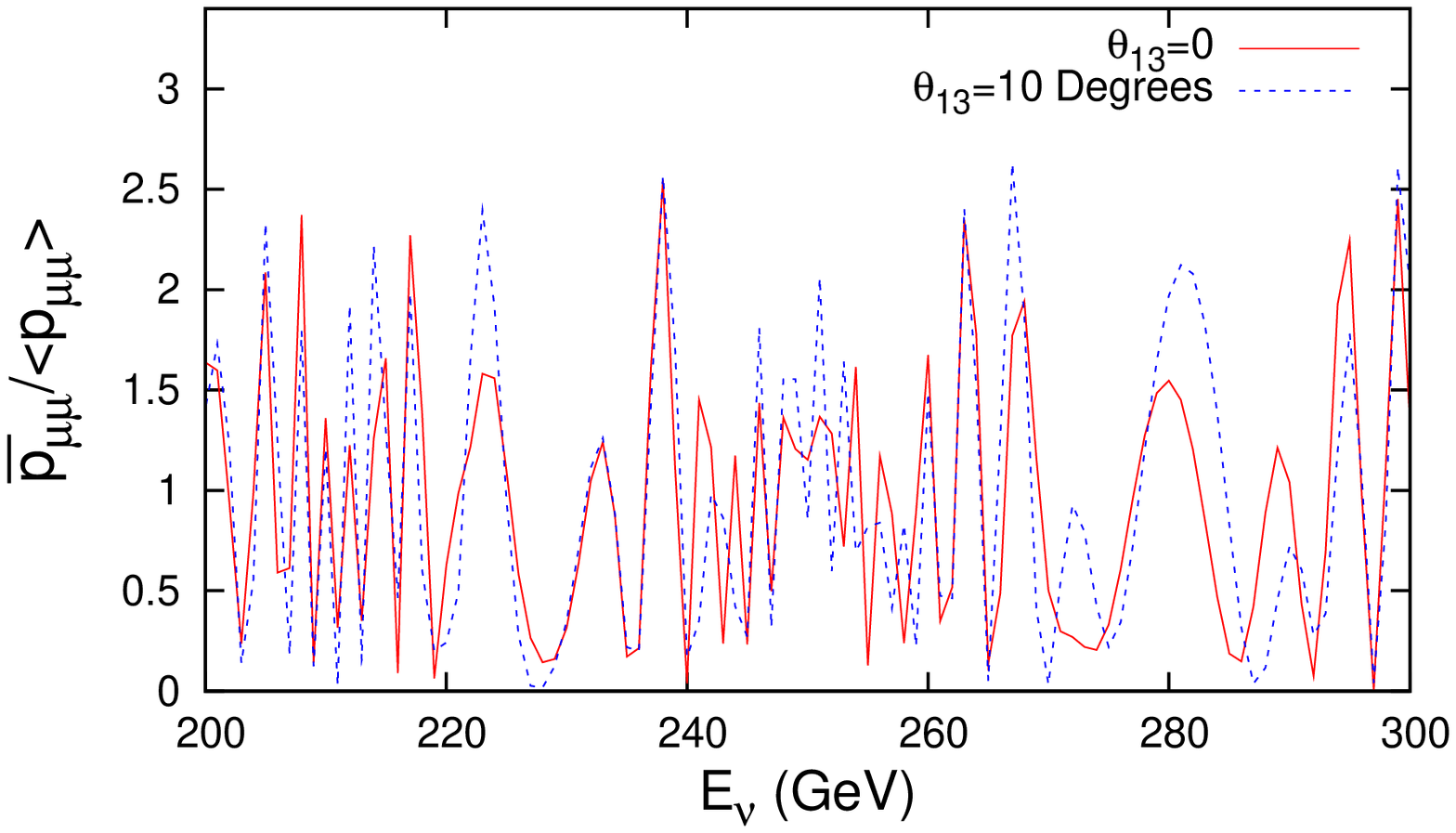}}
 \centerline{\hspace{0.45cm}(a)\hspace{8.45cm}(b)}
 \centerline{\vspace{-2.0cm}}
 \end{center}
 \caption{{\small The dependence of the ratios $\bar{P}_{e\mu}/\langle
P_{e\mu}\rangle$ and $ \bar{P}_{\mu\mu}/\langle P_{\mu\mu}\rangle$
on the energy of the produced neutrinos in the annihilation of the
DM particles inside the Sun. $\bar{P}_{\alpha\beta}$ is the
oscillation probability $\nu_\alpha\to\nu_\beta$
 obtained by integrating over the production
point of neutrinos inside the Sun taking into account the
oscillatory terms. To perform this analysis, we have taken the
neutrino mass scheme to be normal hierarchical. The $\langle
P_{\alpha\beta}\rangle$ is the same quantity without taking into
account the oscillatory terms. Deviation of  the ratio
$\bar{P}_{\alpha\beta}/\langle P_{\alpha\beta}\rangle$ from one is
a measure of the significance of the oscillatory terms. } }
  \label{fig2}
\end{figure}
Let us define $\bar{P}_{\alpha \beta}$ as the average of the
oscillation probability over production point with keeping the
oscillatory terms. Moreover, let us define $\langle {P}_{\alpha
\beta}\rangle$ as the same quantity with dropping the oscillatory
terms. If the ratio $ \bar{P}_{\alpha \beta}/\langle P_{\alpha
\beta}\rangle$ equals to one, it means when we integrate over the
production point, the oscillatory terms average to zero. We have
numerically calculated $ \bar{P}_{\alpha \beta}/\langle P_{\alpha
\beta}\rangle$ and displayed it in Fig.~\ref{fig2}. As seen in
Fig.~\ref{fig2}, $ \bar{P}_{e \mu}/\langle P_{e \mu}\rangle$ and $
\bar{P}_{\mu \mu}/\langle P_{\mu \mu}\rangle$ can substantially
deviate from one which means dropping the oscillatory terms for
monochromatic spectrum is not justified. Thus, the numerical
calculation demonstrated in Fig.~\ref{fig2} confirms our simple
analysis made above. Another interesting point is the significant sensitivity
to $\theta_{13}$. This is understandable because for
$\theta_{13}>0$, the 1-3 resonance in the Sun can play a role. We
repeated the same numerical analysis for the case that the initial
neutrino state is a coherent combination of the neutrino flavor
states and reached the same conclusion.

\section{Width of the spectrum of neutrinos \label{widthSpec}}

The spectrum of neutrinos from direct annihilation of dark matter
particles at rest will be monochromatic. In reality, however
several thermal and quantum mechanical effects lead to widening of
the spectrum. Let us consider them one by one.

\textit{Thermal widening:} The dark matter particles can fall in
gravitational potential of the Sun and despite the very low cross
section can eventually be thermalized. The temperature in the Sun
center is about 1 keV so the average velocity is given by
$\bar{v}=(3 T/m_{DM})^{1/2}\simeq 60~{\rm km/sec}(100~{\rm
GeV}/m_{DM})^{1/2}$. The widening due to this velocity is \be
\label{width}{\Delta E\over E}\sim {\bar{v}\over c}\sim
10^{-4}~\left({T\over 1.3~{\rm keV}}\right)^{1/2}\left({100~{\rm
GeV}\over m_{DM}}\right)^{1/2}. \ee

\textit{Decoherence due to gravitational acceleration:} The DM
particles become accelerated in the gravitational field of the Sun
so neutrinos from the annihilation of a DM pair will have a
coherence length smaller than they would if the acceleration did
not exist. We follow the same logic as in \cite{Farzan:2008eg} to
calculate the wavepacket size. Notice that this is a quantum
mechanical effect originating from the uncertainty principle. Let
us hypothetically divide the path of the DM particles to
successive segments of size $v \Delta \tau$. At each segment, the
wavepacket of the emitted neutrinos will have a width of $\Delta p
\sim 1/\Delta \tau$. During $\Delta \tau$, the velocity of DM
particles is changed by $4\pi G \rho r \Delta \tau/3$ so the
average momentum of the neutrino wavepackets emitted during the
successive segments will differ by $\delta p\sim 4\pi G \rho r
\Delta \tau m_{DM}/3 $. Now, if $\delta p \ll 1/\Delta \tau$, the
two successive wavepackets will form a single wavepacket with
length $2 \Delta \tau$. However, for $\delta p > 1/\Delta \tau$,
the two successive wavepackets are incoherent. That is the
coherence length can be found by equating $\Delta \tau = 1/\delta
p$ which leads to
\be {\Delta p\over p} =\left({4 \pi G \rho r_{DM}\over 3
m_{DM}}\right)^{1/2}\sim 5 \times 10^{-17}\left({100~{\rm GeV}
\over m_{DM}}\right)^{1/2}. \ee Thus, widening of the spectrum due
to acceleration is quite negligible. The width of the spectrum of
the neutrino flux is dominated by thermal fluctuations rather than
quantum mechanical widening.

\textit{Natural width of neutrino wavepacket:} The annihilation
time sets  a natural limit on the wavepacket size of the produced
neutrinos. That is
$$ \frac{\Delta p}{p}> {n_{DM} \langle \sigma_{ann} v \rangle
\over m_{DM}}\ , $$ where $n_{DM}$ is the density of dark matter
in the Sun center. $\langle \sigma_{ann} v\rangle$ is determined
by the DM abundance in the universe: $\langle \sigma_{ann}
v\rangle\sim 10^{-36}~ {\rm cm}^{2}$. Evaluating $n_{DM}$ is more
model-dependent. $n_{DM}$ cannot be larger than $C \tau_\odot /(4
\pi r_{DM}^3/3)$ where $C$ is the capture rate of DM particles by
the Sun and $\tau_\odot$ is the Sun lifetime. $r_{DM}$ determines
the size of volume around the Sun center where the DM density is
relatively high (\textit{see}, Eq.~(\ref{rDM})). Annihilation of
course reduces the DM number density but let us take $n_{DM}\sim
C\tau_\odot/(4\pi r_{DM}^3/3)$ to obtain a conservative estimate
for the natural width. The capture rate is given by \cite{bist}
\be \label{C} C\sim {\rho_{DM} \over m_{DM} v_{DM}}
\left(\frac{M_\odot}{m_p}\right)\sigma_{DM-nucleon} \langle
v_{esc}^2\rangle\ , \ee where $\rho_{DM}=0.39~{\rm GeV}~{\rm
cm}^{-3}$ \cite{Catena} and $v_{DM}\sim 270~{\rm km~sec}^{-1} $
\cite{Kamionkowski:1997xg} are respectively the local density and
velocity of DM particles in our galaxy; $M_\odot$ and $m_p$ are
respectively the Sun and proton masses. Of course,
$\sigma_{DM-nucleon}$ is unknown but if the interaction is
spin-dependent, it can be as high as $\mathcal{O}({\mathrm {pb}})$
\cite{picasso}. The maximal possible capture rate is therefore
$O[10^{24}~{\rm sec}^{-1}]$. Inserting the numerical values, we
find that
$$ {n_{DM}\langle v\sigma_{ann} \rangle \over m_{DM}} \sim
10^{-38}$$ so even with overestimating $n_{DM}\langle
v\sigma_{ann} \rangle$, the natural lower bound is too weak and
$\Delta p/p$ will be dominated by thermal widening; {\textit
i.e.,} Eq.~(\ref{width}).

\textit{Widening due to scattering:} At energies higher than
100~GeV, some of the produced neutrinos can undergo scattering
before leaving the Sun. Neutrinos undergoing charged current
interactions convert into charged leptons which are absorbed in
the matter and do not contribute to the neutrino flux. An
exception is of course $\nu_\tau \to \tau$ because the subsequent
decay of the tau lepton regenerates high energy $\nu_\tau$. On the
other hand, neutrinos undergoing neutral current interactions are
converted to another neutrino with somewhat lower energy. Thus,
the neutrino spectrum emerging from the Sun surface is composed of
a sharp line at $E_\nu =m_{DM}$ superimposed over a continuous
spectrum with $E_\nu < m_{DM}$. The ratio of neutrinos with $E_\nu
\simeq m_{DM}$ to those with $E_\nu<m_{DM}$ depends on the neutral
current cross section which itself depends on the energy of
neutrinos before scattering.  This energy is  in turn  determined
by $m_{DM}$. The mean free path of the neutral current interaction
of the neutrinos with energy 100~GeV in the Sun center is
$$\ell_{NC}=\frac{1}{n_0 \sigma_{NC}}=1.5 \times 10^6~{\rm km} \left({5\times
10^{25}~{\rm cm}^{-3}\over n_0}\right)\left({1.3\times
10^{-37}~{\rm cm}^2 \over \sigma_{NC}}\right)\ ,$$ where we have
used the data from \cite{Bahcall}. Considering the fact that the
matter density in the Sun falls with radius as
$e^{-r/(0.1R_\odot)}$, the ratio of neutrinos undergoing neutral
current  interactions should be of the order of $0.1
R_\odot/\ell_{NC}\simeq 5 \%$. The ratio increases with energy and
at $E_\nu=500$~GeV reaches 35~\%. We restrict our analysis to the
case $m_{DM}<500$~GeV  which is also theoretically and
phenomenologically motivated. In this range, the sharp line in the
spectrum is quite pronounced so we will consider only this sharp
line and will neglect the softened part of the spectrum. If
$m_{DM}$ turns to be greater than about 500~GeV, this analysis
should be reconsidered taking into account the softening effects
due to the neutral current scattering.

\section{Observable
effects of the oscillatory terms: seasonal variation
\label{seasonal}}

For a monochromatic neutrino flux, the rate of $\mu$-track events
in IceCube as a function of time can be estimated as
\begin{align} \frac{d
N_\mu (t)}{dt}&=\int {F_{\nu_\alpha}^0 w P_{\alpha \mu}(t)
(\sigma_{\nu_\mu p}^{CC} \rho_p+ \sigma_{\nu_\mu n}^{CC} \rho_n)
R_\mu A_{eff}(\theta[t])\over [L(t)]^2}\;dV \nonumber \\ &+ \int
{F_{\bar{\nu}_\alpha}^0 \bar{w} P_{\bar\alpha \bar\mu}(t)
(\sigma_{\bar\nu_\mu p}^{CC} \rho_p+\sigma_{\bar\nu_\mu n}^{CC}
\rho_n) R_\mu A_{eff}(\theta[t])\over [L(t)]^2}\;dV
\label{track-rate}
\end{align}
where integration is over the volume inside the Sun where neutrinos
are produced. $F_{\nu_\alpha}^0$ ($F_{\bar{\nu}_\alpha}^0$) is the
flux of $\nu_\alpha$ ($\bar{\nu}_\alpha$) produced in unit volume.
$\rho_p$ and $\rho_n$ are respectively the number densities of
protons and neutrons in the ice. $\sigma_{\nu_\mu p}^{CC}$,
$\sigma_{\nu_\mu n}^{CC}$, $\sigma_{\bar{\nu}_\mu p}^{CC}$ and
$\sigma_{\bar{\nu}_\mu n}^{CC}$ are the cross-sections of the
charged current interactions of $\nu_\mu$ and $\bar{\nu}_\mu$ with
proton and neutron, respectively. $w$ and $\bar{w}$ are suppressions
factors respectively due to the absorption of neutrino and
anti-neutrino fluxes in the Sun. For the energies that we are
interested, $E_\nu
>100~{\rm GeV}$, $w$ and $\bar{w}$ do not depend on the flavor but
in general, $w \ne \bar{w}$. $R_\mu$ is the muon range in the
detector which is the same for muon and anti-muon \cite{range}.
$A_{eff}(\theta[t])$ is the effective area of the detector which
depends on the angle between the neutrino momentum and the axis of
the array of PMTs in detectors, $\theta$. Because of the tilt of
the rotation axis of the Earth, this angle changes as the Earth
moves in its orbit around the Sun. $L(t)$ is the distance between
the Sun and the Earth which varies about $3~\%$ during a year.
Finally, $P_{\alpha \mu}$ and $P_{\bar{\alpha} \bar{\mu}}$ are
respectively the oscillation probability of $\nu_\alpha \to
\nu_\mu$ and $\bar\nu_{{\alpha}}\to\bar {\nu}_\mu$.\footnote{As
discussed in the beginning of section \ref{osc}, $\nu_\alpha$
($\bar{\nu}_\alpha$) can be a coherent combination of different
neutrino flavor eigenstates that diagonalize the density matrix.
We perform our numerical analysis for the case that $\nu_\alpha$
corresponds to a given flavor. However, as discussed, our results
apply to the general case, too. } These probabilities can be
numerically derived from the evolution of neutrino states: \be i
\frac{ d|\nu_\gamma \rangle}{dt}= [{m_\nu^\dagger\cdot m_\nu \over
2 p}+{\rm diag}(V_e,0,0)]_{\gamma \sigma}|\nu_\sigma \rangle
\label{evolution}\ee {\rm and} \be i \frac{ d|\bar\nu_\gamma
\rangle}{dt}= [{m_\nu^T\cdot m_\nu^* \over 2 p}-{\rm
diag}(V_e,0,0)]_{\gamma \sigma}|\bar\nu_\sigma \rangle
\label{evolutionANTI}\ee where $V_e =\sqrt{2} G_F N_e$.

The following remarks are in order:
\begin{itemize}
\item In
Eq.~(\ref{track-rate}), we have neglected the subdominant
contribution from $\nu_\tau\to \tau \to \mu$ which is suppressed
by ${\rm Br}(\tau \to \mu \nu \bar\nu)=17~\%$.
\item
In the case that the spectrum is continuous, the total flux has to
be replaced with differential flux and an integration over energy
has to be taken.
\item Due to the tilt of the Earth rotation axis, during Autumn and Winter
in southern hemisphere, neutrinos entering the IceCube pass
through the mantle of the Earth before reaching the detectors.
However, we can neglect the oscillation of the neutrino inside the
Earth mantle. The reason is that $\Delta m_{ij}^2/p\ll V_e$ so the
effective mixing inside the Earth vanishes and no oscillation
takes place in the constant density of the Earth. (Notice that
although $V_e$ in the Sun center where neutrinos originate is much
larger than $V_e$ inside the Earth mantle, $V_e$ in the Sun
surface is smaller than that in the Earth mantle. That is while
the density in the Earth mantle is almost constant. Because of the
difference in profile, the matter effects in the Sun and Earth are
different.)
\item Production can be either lepton number conserving ({\it
i.e.,} ${\rm DM+DM}\to \nu +\bar{\nu}$) or lepton number violating
({\it i.e.,} ${\rm DM+DM}\to \bar\nu +\bar{\nu}$ or ${\rm
DM+DM}\to \nu +{\nu}$). In the former case, the flux of neutrino
and antineutrino will be obviously the same. In the latter case,
as long as the CP is conserved in annihilation, processes ${\rm
DM+DM}\to \bar\nu +\bar{\nu}$ and ${\rm DM+DM}\to \nu +{\nu}$ take
place with the same rate. Even if CP is violated, any asymmetry in
the $\nu$ and $\bar{\nu}$ fluxes will be a subdominant effect
resulting from the interference of the tree level and loop level
contributions. Thus, we can safely take the initial flux of
neutrinos and antineutrinos to be the same.
\end{itemize}

Let us define \be \label{Ntilde} \tilde{N}(t_0,\Delta t)\equiv
{\int_{t_0}^{t_0+\Delta t} ({d N_\mu }/{dt}) ~dt \over
\int_{t_0}^{t_0+\Delta t} A_{eff}(\theta[t])/[L(t)]^2~dt}. \ee
Notice that if the oscillatory terms average to zero,
$\tilde{N}(t_0,\Delta t)$ will be constant and independent of
$t_0$ and $\Delta t$. Let us therefore define \be \label{delta}
\Delta(t_0,\Delta t)\equiv {\tilde{N}(t_0,\Delta t)-
\tilde{N}(t_0+\Delta t,1~{\rm year}-\Delta t)\over
\tilde{N}(t_0,\Delta t)+ \tilde{N}(t_0+\Delta t,1~{\rm
year}-\Delta t)} \ . \ee In the absence of the oscillatory terms,
$\Delta(t_0,\Delta t)$ vanishes. Deviation of $\Delta$ from zero
is a measure of the strength of the oscillatory terms. For a
continuous spectrum, $\Delta$ vanishes because the oscillatory
terms average to zero so $\Delta$ determines if there is a sharp
feature in the spectrum.

\begin{table}[t]
\begin{center}
\begin{tabular}{|c|c|c|c|c|c|c|c|c|}
\cline{1-9}
& \multicolumn{4}{|c|}{$\Delta ({\rm 20 March, 186 days})$} & \multicolumn{4}{|c|}{$\Delta ({\rm 3 April, 186 days})$}\\
\cline{2-9}
$E_\nu$ (GeV)& \multicolumn{2}{|c|}{$\theta_{13}=0\,^{\circ}$} & \multicolumn{2}{|c|}{$\theta_{13}=10\,^{\circ}$} & \multicolumn{2}{|c|}{$\theta_{13}=0\,^{\circ}$} & \multicolumn{2}{|c|}{$\theta_{13}=10\,^{\circ}$} \\
\cline{2-9}
& NH & IH & NH & IH & NH & IH & NH & IH\\
\hline
100 & 18 \% & 18 \% & 9 \% & 11 \% & 12 \% & 12 \% & 6 \% & 7 \% \\
\hline
300 & 57 \% & 57 \% & 37 \% & 42 \% & 60 \% & 60 \% & 39 \% & 43 \% \\
\hline
\end{tabular}
\caption{\label{eatsource} The values of $\Delta(t_0,\Delta t)$
defined in Eq.~(\ref{delta}) for both normal and inverted
hierarchies of the neutrino's mass spectrum. The initial flux is
taken to be composed of $\nu_e$ and $\bar{\nu}_e$ with two
different values of energies $E_\nu=100$ and 300~GeV. On $t_0=3$
April, the earth reaches the point on its orbit depicted in
Fig.~\ref{90and270} and $t_0=20$ March corresponds to the spring
equinox.} \vspace{1cm}
\end{center}
\end{table}

\begin{table}[t]
\begin{center}
\begin{tabular}{|c|c|c|c|c|c|c|c|c|}
\cline{1-9}
& \multicolumn{4}{|c|}{$\Delta ({\rm 20 March, 186 days})$} & \multicolumn{4}{|c|}{$\Delta ({\rm 3 April, 186 days})$}\\
\cline{2-9}
$E_\nu$ (GeV)& \multicolumn{2}{|c|}{$\theta_{13}=0\,^{\circ}$} & \multicolumn{2}{|c|}{$\theta_{13}=10\,^{\circ}$} & \multicolumn{2}{|c|}{$\theta_{13}=0\,^{\circ}$} & \multicolumn{2}{|c|}{$\theta_{13}=10\,^{\circ}$} \\
\cline{2-9}
& NH & IH & NH & IH & NH & IH & NH & IH\\
\hline
100 & 9 \% & 6 \% & 4 \% & 1 \% & 7 \% & 4 \% & 3 \% & 0.3 \% \\
\hline
300 & 12 \% & 7 \% & 6 \% & 19 \% & 13 \% & 7 \% & 6 \% & 20 \% \\
\hline
\end{tabular}\caption{\label{muatsource} The same as Table \ref{eatsource}
except that the initial flux is taken to be composed of $\nu_\mu$
and $\bar{\nu}_\mu$.} \vspace{0.5cm}
\end{center}
\end{table}

\begin{figure}[t!]
  \begin{center}
 \centerline{\includegraphics[bb=350 100 500
 500,keepaspectratio=true,clip=true,angle=-90,scale=0.9]{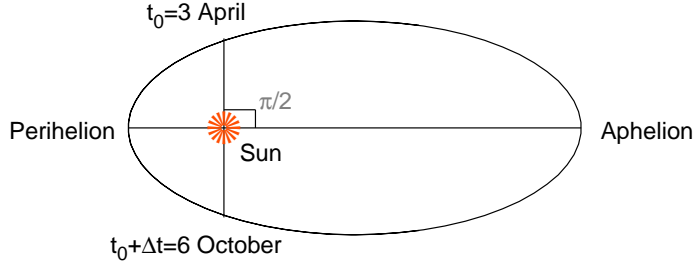}}
 \centerline{\vspace{-2.0cm}}
 \end{center}
 \caption{{\small Positions of the Earth on $t_0=3$ April
 and $t_0+\Delta t=6$ October, where $\Delta t=186$ days.  } }
  \label{90and270}
\end{figure}

Tables~\ref{eatsource}~and~\ref{muatsource} show values of
$\Delta$ for different values of $m_{DM}(=E_\nu)$, $\theta_{13}$,
$(t_0,\Delta t)$ and for both Normal Hierarchical (NH) and
Inverted Hierarchical (IH) neutrino mass schemes. To fill the
Table~\ref{eatsource}, we have taken the initial flux at source to
be composed of merely $\nu_e$ and $\bar{\nu}_e$ and in the case of
Table~\ref{muatsource}, we have taken the initial flux to be
composed of only $\nu_\mu$ and $\bar{\nu}_\mu$. For one set of the
data, we have chosen ($t_0=20$ of March, $\Delta t=186$ days).
Twentieth of March corresponds to spring equinox\footnote{Spring
equinox is called ``Norooz'' in Farsi and is the beginning of the
year in the Iranian calendar.} and $t_0+\Delta t$ corresponds to
autumn equinox. For another set, we have chosen $t_0=3$ of April
and $\Delta t=186$ days which correspond to the points in the
orbit indicated in Fig.~\ref{90and270}.


We have taken two values of $\theta_{13}$ ($\theta_{13}=0$  and
$\theta_{13}=10^\circ$) and set the CP-violating phase, $\delta$,
equal to zero. We have calculated $\Delta (t_0,\Delta t)$ for the
case of exactly monochromatic spectrum as well as for a narrow
Gaussian with width given by the thermal fluctuations of the dark
matter [{\it see}, Eq.~(\ref{width})]. As expected from the
previous section, the difference in $\Delta$ for two cases
(monochromatic versus Gaussian with the width given in
Eq.~(\ref{width})) is quite negligible. In both cases, $\Delta$
significantly deviates from zero which means seasonal variation
due to oscillatory terms is significant and potentially
measureable.

Several features are obvious in the figures of
Tables~\ref{eatsource}~and~\ref{muatsource}. As can be seen from
Table~\ref{eatsource}, in the case $\theta_{13}=0$, the values of
$\Delta (t_0,\Delta t)$ (for both $t_0=20$ March and $t_0=3$
April) are equal for NH and IH. This equality is a consequence of
the fact that for $\theta_{13}=0$, the contribution of $\nu_3$ to
$\nu_e$ is zero so both the NH and IH cases are equivalent. By
comparing the figures in Table~\ref{eatsource} with the figures of
Table~\ref{muatsource} it can be seen that the value of $\Delta
(t_0,\Delta t)$ is typically larger for the annihilation of DM
particle to electron neutrinos (DM+DM$\to\nu_e\bar{\nu}_e$). Thus,
if the measurements of neutrino telescopes (such as IceCube) show
a large value for $\Delta (t_0,\Delta t)$, the annihilation of DM
particles to $\nu_e$ (and $\bar{\nu}_e$) will be favored.

{In the real experiments such as IceCube, measurement of $\Delta
(t_0,\Delta t)$ will be more tricky and statistical errors and
background events have to be taken into account carefully. The
statistical error of the ratio $\Delta(t_0,\Delta t)$ in
Eq.~(\ref{delta}) is given by the following formula}
\begin{align}\label{stat-error}
\delta\Delta(t_0,\Delta t)={2\left(\tilde{N}(t_0,\Delta
t)\tilde{N}(t_0+\Delta t,1 {\text {year}}-\Delta
t)\right)^{1/2}\over \left(\tilde{N}(t_0,\Delta
t)+\tilde{N}(t_0+\Delta t,1 {\text {year}}-\Delta t)\right)^{3/2}}
,\end{align} {where we have inserted $\delta
\tilde{N}=\sqrt{\tilde{N}}$.
To evaluate $\delta \Delta$, we should estimate what is the
maximum number of events per year allowed within the present
bounds. The bound from present observation on the total number of
muon tracks depends on the shape of spectrum of the neutrinos.
This is understandable because the detection threshold of AMANDA
and its augmented version IceCube are not exactly the same,
especially once DeepCore is added. In \cite{amanda-sun}, an
analysis has been made for the spectrum corresponding to
annihilation into $W^+W^-$ as well as into $\tau^+\tau^-$ for
various values of the DM mass. The result is that if the bound is
saturated, depending on the DM mass or the shape of the spectrum,
a ${\rm km}^3$ scale detector can observe between 500 to a few
1000 events per year. The spectrum for annihilation into neutrino
pair is harder so the bound will be stronger. Detailed analysis is
beyond the scope of the present paper. Taking total number of 400
/year/km$^2$ however sounds reasonable. With 400 muon tracks, the
statistical error, $\delta \Delta$, from Eq.~(\ref{stat-error}),
is less than $0.05$. From Eq.~(\ref{C}), we find that this
corresponds to nucleon DM cross section of $10^{-4}$pb which is
well below the bound on spin-dependent cross section. Typical
values for $\Delta$ in Tables~\ref{eatsource}~and~\ref{muatsource}
can therefore be established after a few years of data taking
({\it i.e.,} $\delta \Delta \ll \Delta$). In practice, the
measurement will suffer from reducible and irreducible
backgrounds. A well-known irreducible background comes from the
``solar atmospheric neutrinos''. That is neutrinos produced by
interaction of cosmic rays with the atmosphere of the Sun which
amounts to about 10 events per year \cite{solar-atmosphere}. The
signal can be larger than this background by one order of
magnitude. Moreover, the solar atmospheric flux, its energy
spectrum as well as its flavor composition can be calculated so
its effects can be subtracted.
%
Two other sources of backgrounds exist for the $\mu$-track events
from the Sun: the atmospheric neutrinos and the atmospheric muons.
From the spring to autumn equinoxes the Sun is below the horizon
at the IceCube site and the neutrinos coming from the Sun produce
up-ward going $\mu$-tracks. The atmospheric muons will be absorbed
in this period of time by the surrounding material of the
detector. This results in huge suppression of atmospheric muon
background such that it can be completely neglected.
During the autumn to spring equinoxes, the Sun is above the
horizon for IceCube and the number of atmospheric muons increases
drastically. However, by using the outer parts of the IceCube as a
veto, the atmospheric muon background will be quite low at the
DeepCore \cite{deepcore} (the surrounding area can veto
atmospheric muon events up to one part in $10^6$).
Let us now discuss the muon-track events induced by atmospheric
neutrinos. As shown in appendix C of Ref.\cite{background}, thanks
to the high angular resolution of the IceCube for $\mu$-tracks, by
focusing on a cone with half angle $1^\circ$ around the direction
of Sun, the number of background through-going $\mu$-tracks
induced by atmospheric neutrinos can be reduced to $\sim
6$~yr$^{-1}$ for the whole energy spectrum above the IceCube
energy threshold of the $\mu$-track detection (number of
background contained $\mu$-track events in the DeepCore is $\sim
2.5$~yr$^{-1}$). Thus, the background events from the atmospheric
neutrinos are not also a limiting factor in the measurement of
$\Delta (t_0,\Delta t)$.  }

In the previous section, we observed that $P_{e\mu}$ is quite
sensitive to $\theta_{13}$. The reason is that for
$\theta_{13}=0$, no 1-3 resonance in $\nu_e \to \nu_\mu$ takes
place but for $\theta_{13}>0$, such a resonance plays a
significant role. From the tables, we observe that $\Delta$ is
also quite sensitive to $\theta_{13}$. Considering this high
sensitivity, it is tempting to entertain the possibility of
extracting $\theta_{13}$ from $\Delta$ but absence of knowledge on
the initial flavor composition renders the method useless.

\section{Discussion of results \label{dis}}
Throughout this paper we have focused on a monochromatic spectrum
({\it i.e.}, a very narrow line) of neutrinos from the direct
annihilation of dark matter particles to neutrino pairs inside the
Sun. In reality, as we discussed earlier, the neutral current
interactions inside the Sun will smear a part of the sharp line
into a continuous spectrum with lower energy. Moreover, along with
${\rm DM}+{\rm DM}\to \stackrel{(-)}{\nu}+ \stackrel{(-)}{\nu}$,
DM particles can annihilate into $\tau \bar\tau$  and other
particles whose subsequent decay will lead to a continuous
spectrum with lower energy. Thus, the spectrum will be composed of
a sharp line superimposed on the upper edge of a continuous
background. Our discussion holds valid about the sharp line part
of the spectrum.

In neutrino telescopes such as IceCube, the direction of
$\mu$-track can be reconstructed by amazing precision of $1^\circ$
\cite{icecube} which means neutrinos from the Sun can be singled
out. In practice, the measurement of the spectrum is going to be
challenging especially that a large fraction of muons are produced
by interaction of $\nu_\mu$ outside the detector and lose energy
before entering the detector. However, if the statistics is high
enough, the spectrum can be reconstructed by measuring the energy
of contained muons. Let us consider different situations one by
one.

In case that the spectrum is not reconstructed, the measurement of
$\Delta$ gives invaluable information on the shape of the
spectrum. From Tables~\ref{eatsource}~and~\ref{muatsource}, we see
that $\Delta$ can be quite sizeable for both normal and inverted
hierarchical mass schemes. Thus, if the number of events is a few
hundreds, deviation of $\Delta$ from zero can be established.
Since we have divided the number of events by $\int L^{-2} A_{eff}
dt$ in the definition of $\tilde{N}(t_0,\Delta t)$, a deviation of
$\Delta$ from zero indicates that the oscillatory terms in
oscillation probability do not average to zero ({\it see,}
Eqs.~(\ref{Ntilde},\ref{delta})). This in turn shows that there
must be sharp features in the spectrum, originating from direct
annihilation of DM pairs to neutrino pairs. To reach such a
conclusion, the possibility of other seasonal modulation (like
detector performance or seasonal variation of the background
\cite{hep-ex}) has to be of course subtracted.

In case that the spectrum is reconstructed, $\Delta$ again
provides invaluable information. If the spectrum contains a sharp
line, we in general expect $\Delta$ to be nonzero. If a sharp line
is observed in the spectrum but $\Delta$ turns out to be zero, the
most natural explanation is that the initial flavor ratio is
$F^0_{\nu_e}=F^0_{\nu_\mu}=F^0_{\nu_\tau}$ which in turn means the
physics of DM annihilation is flavor blind. This may be a unique
way to learn about flavor composition as the cascade events most
probably will lie below the detection threshold for
$m_{DM}\lesssim500$ GeV. In general, to analyze the properties of
dark matter, seasonal variation, $\Delta$, provides a powerful
tool.

\section{Concluding remarks \label{Con}}

In case that DM pairs annihilate into neutrino pairs, there will
be a sharp line in the spectrum. We have shown that in the
presence of such a line, the oscillatory terms in the oscillation
probability do not average to zero and can therefore lead to a
seasonal variation of the number of events in neutrino detectors.
We have shown that the main cause for widening of the line is the
thermal velocity distribution of DM particles inside the Sun but
the widening will be too small to lead to vanishing of the
oscillatory effects. We have demonstrated that the uncertainty in
the production point cannot lead to vanishing of the oscillatory
effects, either.

We have defined an observable quantity, $\Delta$,  whose deviation
from zero is a measure of the significance of the oscillatory
terms ({\it see,} Eq.~(\ref{delta})). We have shown that $\Delta$
can reach as high as 60~\% so its deviation from zero can be
established by a few hundred muon-track events. We have calculated
the background and statistical error and have found that in the
case that the flux is close to the present bounds, measuring
$\Delta$ is doable. $\Delta$ contains invaluable information on
the properties of DM particles. If $\Delta \ne 0$, even without
performing the challenging energy spectrum reconstruction, we may
conclude that there is a sharp line in the spectrum so the DM
pairs have an annihilation mode into neutrino pairs. If the
spectrum is reconstructed and a sharp line is identified but
$\Delta$ turns out to be zero, a natural explanation is that at
the source all three flavors are produced by equal amounts. This
in turn means that DM annihilation is flavor blind. Considering
that for low values of $E_\nu$ ({\it i.e.,} for $m_{DM}\lesssim
500$~GeV), the cascade events will be below detection threshold of
neutrino telescopes, $\Delta$ might indirectly provide a unique
probe of flavor composition.

As discussed, the width of the line is given by thermal
fluctuations in the Sun center ($r<r_{DM}<0.01 R_\odot$). Even
conventional solar neutrinos from thermonuclear processes are
produced mainly in the outer layers so our information on this
region depends merely on solar models. Our results are however
robust against solar models as $\Delta p/p$ scales as $T^{1/2}$
({\it see,} Eq.~(\ref{width})). In order for the oscillatory terms
given by $\Delta m_{12}^2$ to be erased, $T$ has to be 10000 times
larger than the value predicted by solar models which seems quite
unlikely.

\section*{Acknowledgement} We are grateful to A. Yu Smirnov,
R. Allahverdi, P.~Serpico and P.~Gondolo for useful comments. A.
E. appreciates ``Bonyad-e-Melli-e-Nokhbegan'' for partial
financial support.


\begin{thebibliography}{10}



\bibitem{pdg}
  C.~Amsler {\it et al.}  [Particle Data Group],
  Phys.\ Lett.\  B {\bf 667} (2008) 1.


\bibitem{Hooper:2009zm}
  D.~Hooper,
  arXiv:0901.4090 [hep-ph];
  F.~D.~Steffen,
  Eur.\ Phys.\ J.\  C {\bf 59} (2009) 557
  [arXiv:0811.3347 [hep-ph]];
  M.~E.~Peskin,
  J.\ Phys.\ Soc.\ Jap.\  {\bf 76} (2007) 111017
  [arXiv:0707.1536 [hep-ph]];
  H.~Murayama,
  arXiv:0704.2276 [hep-ph];
  D.~Hooper and S.~Profumo,
  Phys.\ Rept.\  {\bf 453} (2007) 29
  [arXiv:hep-ph/0701197].








\bibitem{WIMP}
  G.~Jungman, M.~Kamionkowski and K.~Griest,
  Phys.\ Rept.\  {\bf 267} (1996) 195
  [arXiv:hep-ph/9506380];
  G.~Bertone, D.~Hooper and J.~Silk,
  Phys.\ Rept.\  {\bf 405} (2005) 279
  [arXiv:hep-ph/0404175].


\bibitem{suncapture}
  W.~H.~Press and D.~N.~Spergel,
  Astrophys.\ J.\  {\bf 296} (1985) 679;


\bibitem{sunneutrino}
  G.~B.~Gelmini, P.~Gondolo and E.~Roulet,
  Nucl.\ Phys.\  B {\bf 351} (1991) 623.
  M.~Kamionkowski,
  Phys.\ Rev.\  D {\bf 44} (1991) 3021;
  S.~Ritz and D.~Seckel,
  Nucl.\ Phys.\  B {\bf 304} (1988) 877.
  V.~D.~Barger, F.~Halzen, D.~Hooper and C.~Kao,
  Phys.\ Rev.\  D {\bf 65} (2002) 075022
  [arXiv:hep-ph/0105182];
  V.~Barger, W.~Y.~Keung, G.~Shaughnessy and A.~Tregre,
  Phys.\ Rev.\  D {\bf 76} (2007) 095008
  [arXiv:0708.1325 [hep-ph]];
  A.~de Gouvea,
  Phys.\ Rev.\  D {\bf 63} (2001) 093003
  [arXiv:hep-ph/0006157];



\bibitem{gribov}
V.~N.~Gribov and B.~Pontecorvo,
  Phys.\ Lett.\  B {\bf 28} (1969) 493.



\bibitem{weiler}
R.~Lehnert and T.~J.~Weiler,
  Phys.\ Rev.\  D {\bf 77} (2008) 125004
  [arXiv:0708.1035 [hep-ph]];
  A.~E.~Erkoca, M.~H.~Reno and I.~Sarcevic,
  Phys.\ Rev.\  D {\bf 80} (2009) 043514
  [arXiv:0906.4364 [hep-ph]];
  M.~Cirelli, N.~Fornengo, T.~Montaruli, I.~Sokalski, A.~Strumia and
  F.~Vissani,
  Nucl.\ Phys.\  B {\bf 727} (2005) 99
  [Erratum-ibid.\  B {\bf 790} (2008) 338]
  [arXiv:hep-ph/0506298].



\bibitem{mono}
  V.~D.~Barger, W.~Y.~Keung and G.~Shaughnessy,
  Phys.\ Lett.\  B {\bf 664} (2008) 190
  [arXiv:0709.3301 [astro-ph]].


\bibitem{mono1}
M.~Blennow, H.~Melbeus and T.~Ohlsson,
  arXiv:0910.1588 [hep-ph];



\bibitem{Blennow:2007tw}
  M.~Blennow, J.~Edsjo and T.~Ohlsson,
  JCAP {\bf 0801} (2008) 021
  [arXiv:0709.3898 [hep-ph]].




\bibitem{Farzan:2009ji}

  C.~Boehm, Y.~Farzan, T.~Hambye, S.~Palomares-Ruiz and S.~Pascoli,
  Phys.\ Rev.\  D {\bf 77} (2008) 043516
  [arXiv:hep-ph/0612228].

  \bibitem{models} Y.~Farzan,
  Phys.\ Rev.\  D {\bf 80} (2009) 073009
  [arXiv:0908.3729 [hep-ph]];
Y.~Farzan, S.~Pascoli and M.~A.~Schmidt,
  arXiv:1005.5323 [hep-ph];
 Y. Farzan, \textit{work in progress}.



\bibitem{Allahverdi}
  R.~Allahverdi, S.~Bornhauser, B.~Dutta and K.~Richardson-McDaniel,
  Phys.\ Rev.\  D {\bf 80} (2009) 055026
  [arXiv:0907.1486 [hep-ph]].
  If the masses of the right-handed neutrinos and sneutrinos are
  degenerate ({\it i.e.,} the supersymmetric limit), the neutrino
  flux will emerge to be monochromatic.






\bibitem{Farzan:2008eg}
  Y.~Farzan and A.~Y.~Smirnov,
  Nucl.\ Phys.\  B {\bf 805} (2008) 356
  [arXiv:0803.0495 [hep-ph]].




\bibitem{bist}
  A.~Gould,
  Astrophys.\ J.\  {\bf 328} (1988) 919;
A.~Gould,
  Astrophys.\ J.\  {\bf 321} (1987) 571.
\bibitem{Catena}
  R.~Catena and P.~Ullio,
  arXiv:0907.0018 [astro-ph.CO].

\bibitem{Kamionkowski:1997xg}
  M.~Kamionkowski and A.~Kinkhabwala,
  Phys.\ Rev.\  D {\bf 57} (1998) 3256
  [arXiv:hep-ph/9710337].




\bibitem{picasso}
S.~Archambault {\it et al.},
  Phys.\ Lett.\  B {\bf 682} (2009) 185
  [arXiv:0907.0307 [hep-ex]].


\bibitem{Bahcall}
  J.~N.~Bahcall, M.~H.~Pinsonneault and S.~Basu,
  Astrophys.\ J.\  {\bf 555} (2001) 990
  [arXiv:astro-ph/0010346];
  R.~Gandhi, C.~Quigg, M.~H.~Reno and I.~Sarcevic,
  Astropart.\ Phys.\  {\bf 5} (1996) 81
  [arXiv:hep-ph/9512364].


\bibitem{range}
  D.~E.~Groom, N.~V.~Mokhov and S.~I.~Striganov,
  Atom.\ Data Nucl.\ Data Tabl.\  {\bf 78} (2001) 183.



\bibitem{amanda-sun}
  J.~Braun, D.~Hubert and f.~t.~I.~Collaboration,
  arXiv:0906.1615 [astro-ph.HE].

\bibitem{solar-atmosphere}
  G.~L.~Fogli, E.~Lisi, A.~Mirizzi, D.~Montanino and P.~D.~Serpico,
  Phys.\ Rev.\  D {\bf 74} (2006) 093004
  [arXiv:hep-ph/0608321].
  G.~L.~Fogli, E.~Lisi, A.~Mirizzi, D.~Montanino and P.~D.~Serpico,
  Nucl.\ Phys.\ Proc.\ Suppl.\  {\bf 168} (2007) 283
  [J.\ Phys.\ Conf.\ Ser.\  {\bf 120} (2008) 052039].



\bibitem{background}
  V.~Barger, J.~Kumar, D.~Marfatia and E.~M.~Sessolo,
  arXiv:1004.4573 [hep-ph].


\bibitem{deepcore}
  C.~Wiebusch and f.~t.~I.~Collaboration,
  arXiv:0907.2263 [astro-ph.IM].




\bibitem{icecube}
  J.~Ahrens {\it et al.}  [IceCube Collaboration],
  Astropart.\ Phys.\  {\bf 20} (2004) 507
  [arXiv:astro-ph/0305196].




\bibitem{hep-ex}

  P.~Adamson {\it et al.}  [MINOS Collaboration],
  Phys.\ Rev.\  D {\bf 81} (2010) 012001
  [arXiv:0909.4012 [hep-ex]].





\end{thebibliography}
\end{document}